**Effect of fluorinated diamond surface on charge state of nitrogen-vacancy centers**

Shanying Cui and Evelyn L. Hu[a]

*School of Engineering and Applied Sciences, Harvard University, Cambridge, MA 02138 USA*

We investigated the effect of fluorine-terminated diamond surface on the charged state of shallow nitrogen vacancy defect centers (NVs). Fluorination is achieved with $CF_4$ plasma and the surface chemistry is confirmed with x-ray photoemission spectroscopy. Photoluminescence of these ensemble NVs reveal that fluorine-treated surfaces lead to a higher negatively charged nitrogen vacancy ($NV^-$) population than oxygen-terminated surfaces. Using surface chemistry to control NV charges, in particular increasing the density of $NV^-$ centers, is an important step towards improving the optical and spin properties of NVs for quantum information processing and magnetic sensing.

The negatively charged nitrogen-vacancy ($NV^-$) center in diamond is a luminescent point defect with applications in quantum information processing, high sensitivity magnetometry, and biotagging.[1] The defect can also exist in a neutral charge state ($NV^0$), but the attractive optical and spin properties have only been observed in $NV^-$. Recently, significant progress has been made in coupling $NV^-$ centers to an external spin[2,3] and increasing photon extraction from the diamond lattice.[4] Both applications require the NVs to be near the surface while maintaining spin coherence and/or optical activity. However, favorable properties, such as $NV^-$ luminescence and ESR linewidth, are compromised in shallow NVs.[5] It is therefore imperative to understand the effect of surface chemistry on the charge state of near-surface NVs and to maximize the number of spin-active $NV^-$ states over spin-inactive $NV^0$.

---

[a] Author to whom correspondence should be addressed. Electronic mail: ehu@seas.harvard.edu

It is well known that hydrogen-termination leads to the depletion of electrons at the surface of the diamond and hence a high $NV^0$ population in shallow NVs.[6–9] Hydrogen is less electronegative than carbon and a hydrogen-terminated diamond surface exhibits a negative electron affinity. Adsorbed water layers on the surface accept electrons and this electron transfer bends the surface bands upwards, resulting in hole accumulation at the surface and a primarily $NV^0$ population. In contrast, oxygen-terminated surfaces lead to a relatively higher $NV^-$ density, which is attributed to the relative electronegativity of oxygen to carbon. The C-O surface dipole points in the opposite direction of the C-H dipole, resulting in a positive electron affinity for the diamond. Oxygen-termination is currently the state-of-the-art surface termination for diamond and other chemical terminations to further increase the $NV^-$ population have not been reported.

Due to the high electronegativity of fluorine, the C-F bond is more polar than C-O. *Ab initio* calculations of fluorinated diamonds show a stable and full coverage and high electron affinity.[10] Recently, the fluorinated diamond electron affinity has been measured experimentally to be 2.56 eV, about 0.43 eV higher than for the oxygen-terminated surface.[11] Various methods have been reported in the literature for fluorine-terminating diamond surfaces, including exposure to $XeF_2$ gas[11,12], $CF_4$ plasma[13], and $CHF_3$ plasma.[14] However, all prior work on diamond fluorination has been confined to surface chemistry, where as this work studies the effect of surface fluorination on the charge states of NV centers in bulk diamond.

Experiments were conducted on electronic grade, CVD-grown, (100)-oriented electronic grade diamond (Element6). Surface characterization was carried out with X-ray photoemission spectroscopy (XPS) and charging of the insulating diamond surface was compensated with low energy electrons and an argon ion beam (Thermo Scientific K-alpha). When the stage is not tilted, the XPS signal comes from core electrons within approximately 6 nm of the surface (three times the mean-free path of electrons in diamond).[15] Photoluminescence (PL) spectra were collected using 532 nm laser excitation in a confocal microscope with a 0.5 μm² spot size. Diamonds were first implanted with $^{14}$N at 10 keV at $10^{12}$ ions/cm² (CORE Systems) and then annealed in vacuum at 800°C for 2 hours to create NVs approximately 15nm ± 5nm deep[7]. To both remove graphitic carbon at the surface and to oxygen-terminate the surface, the samples were cleaned with a boiling solution of 1:1:1 nitric: sulfuric: perchloric acid for 2 hours (boiling triacid). Prior to fluorine-treatment, a photoresist mask (Shipley series) was deposited on half of the sample to protect the oxygen-termination (see schematic in Figure 1). Fluorine-treatment was carried out in a Technics RF plasma reactor at 25 sccm $CF_4$ flow rate, 150 W plasma power, and 300 mTorr chamber pressure for 5 minutes. In all cases of plasma-treated surfaces, care was taken to avoid etching or otherwise damaging the surface by choosing plasma treatments with little or no bias voltages and minimizing the interaction time with the plasma. XPS was used to confirm the process of depositing the photoresist mask and removing the mask with acetone and isopropanol did not affect surface chemistry.

The photoelectron spectra of the masked and $CF_4$ plasma-treated half of the same sample is illustrated in Figure 2. The survey spectrum (top inset) reveals 8.5% oxygen and 91.5% carbon concentration on the masked side, whereas the side exposed to $CF_4$ shows 7.1% fluorine, 2.0% oxygen, and 90.9% carbon composition. The difference in surface chemistry between the two sides of the sample is further highlighted in the high resolution scan of the C 1s core electrons and its deconvoluted peaks. Due to the insulating nature of diamond, charging on the surface can shift the detected binding energy of electrons. The spectrum of the masked side is calibrated by aligning the main peak to the reported value of $sp^3$ C-C bond,[16] shifting the spectrum from the measured peak at 286 eV to 285 eV. A graphitic $sp^2$ carbon peak is visible at 283.6 eV. Two types of oxidized carbon peaks are visible at 286.3 and 287.7 eV, corresponding to C-O and C=O, respectively, evidence that the masked region is oxygen-terminated. In contrast to the oxygen-terminated side, the main peak on the unmasked side cannot be calibrated to the bulk diamond lattice, since the lowest binding energy peak is 2 eV lower than the main peak and the $sp^2$ C-C bond has only been reported to be between 0.6-1.5 eV lower than the $sp^3$ C-C.[14] The peak with lowest binding energy has been calibrated from the measured 285.4 eV to the reported 285 eV value. The main peak at 286.9 eV mostly likely reflects a C-CF bond[13,17] and the peaks at 288.5 eV and 290.3 eV (better seen in inset) correspond well to C-F and $C-F_2$ literature values[17,18]. The high concentration of C-CF bonds and the presence of $C-F_2$ suggests some fluorocarbon deposition from the $CF_4$ plasma, which also has been reported previously in literature[13]. We estimate the polymerized fluorocarbon layer to be about 3nm thick, given that the C-C bond disappears at a 60 degree stage tilt, which reduces the analysis depth by half.

The effect of surface termination on NV charge ratio is determined through NV luminescence. A representative PL spectrum in Figure 3(a) shows $NV^0$ and $NV^-$ zero phonon lines (ZPL) at 575 and 638 nm respectively. To estimate the relative probability for the NV center to be in the $NV^-$ state, we analyzed the $NV^0$ and $NV^-$ ZPL individually, by subtracting a linear background around the ZPL and integrating under the peak. Figure 3(b) highlights the ratio of the $NV^-$ ZPL area to the total NV ZPL area of the same sample over the course of different treatments. Throughout the process, there was no change in overall NV intensity, only the ratio of $NV^-$ and $NV^0$ peaks. We started with a boiling triacid-cleaned and oxygen-terminated surface, where PL was taken at multiple points over the whole sample. Error bars here represent standard deviations in the calculated ratio between three to nine spots on the sample. After masking and $CF_4$ plasma treatment, a relative increase of $NV^-$ was observed on the fluorine-terminated side. The fluorine-terminated side remained stable after exposure to air for ten days, whereas the NV$^-$ ratio on the oxygen-terminated side was further degraded. The lack of stability of the oxygen-terminated surface is not well understood and further investigation is required. Oxygen-termination is a mixed chemical state on the surface and can exist as hydroxyl groups, ethers, ketones, or carboxylic acids. Consequently, the oxidation state and the oxygen coverage can change upon reaction with moisture in air. Calculations have shown that diamond terminated with hydroxyl groups also exhibit negative electron affinities,[19] and adsorbed water states are likely to become electron acceptors, as seen with hydrogen-terminated surfaces. The thin polymerized fluorocarbon layer on the $CF_4$ treated side, however, can protect the surface against oxidation and other surface degradations. We returned the whole sample to oxygen-termination by first annealing at 300°C to remove the fluorocarbon layer, then treating with boiling triacid. The NV$^-$ / (NV$^0$ + NV$^-$) ratio on both sides of the sample converged to approximately the initial value.

We compared the effect of hydrogen-, oxygen-, and fluorine-termination without a photoresist mask with a different diamond sample, implanted with $^{14}$N under the same conditions. The sample was first hydrogen-terminated in a $H_2$ plasma by flowing 400 sccm of $H_2$ in a CVD reactor at 60 torr pressure with a 700 W plasma power for 3 minutes. Next, oxygen-termination was achieved with the boiling triacid clean, and finally the sample was fluorine-terminated with $CF_4$ plasma. Between each step, we measured PL at several spots and compared the NV$^-$ and NV$^0$ signals. Our experimental results indicate more electronegative terminations lead to higher relative NV$^-$ concentration (Figure 4). Fluorine-termination induces a downward band-bending,[11] and the strength of the surface dipole can play a significant role in electron density at the surface, thus increasing the relative NV$^-$ signal. Additionally, it is important for the surface termination to have good coverage, since dangling bonds can be electron traps. Previous theoretical calculations show fluorine adsorption energy remains low for all coverages, meaning it is possible to have a full monolayer of F on a diamond <100> surface, whereas full and stable coverage is less likely for other terminations, such as Cl[10].

In summary, we demonstrated that fluorine-termination from $CF_4$ plasma treatments yields a higher $NV^-$ population than oxygen-termination, which is currently used for all magnetometry and photonic diamond devices. From our photoluminescence data on H-, O-, and F-terminated surfaces with nearby NVs, we observe a correlation between the surface dipole strength and the ratio of negatively charged NV centers to neutrally charged NVs. Chemical analysis on the surface reveals a polymerized fluorocarbon on the surface that may stabilize and protect the surface, compared to the chemically mixed oxygen-terminated surfaces. Understanding the effect of surface chemistry on NV signal and NV charges is an important step towards building higher sensitivity magnetometers.

**Acknowledgements**

This work was supported by the Defense Advanced Research Projects Agency QuASAR program, the Center for Nanoscale Systems (CNS) at Harvard University, and the National Science Foundation Graduate Research Fellowship Program.

[19] S. Sque, R. Jones, and P. Briddon, Physical Review B **73**, 1 (2006).

**Figures**

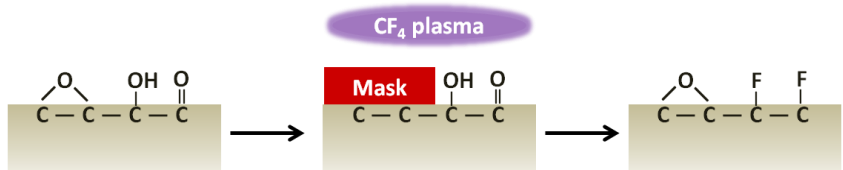

Figure 1: (Color online) Schematic of a diamond sample half oxygen-terminated and half-fluorine-terminated. Oxygen-termination is achieved with a boiling triacid clean.

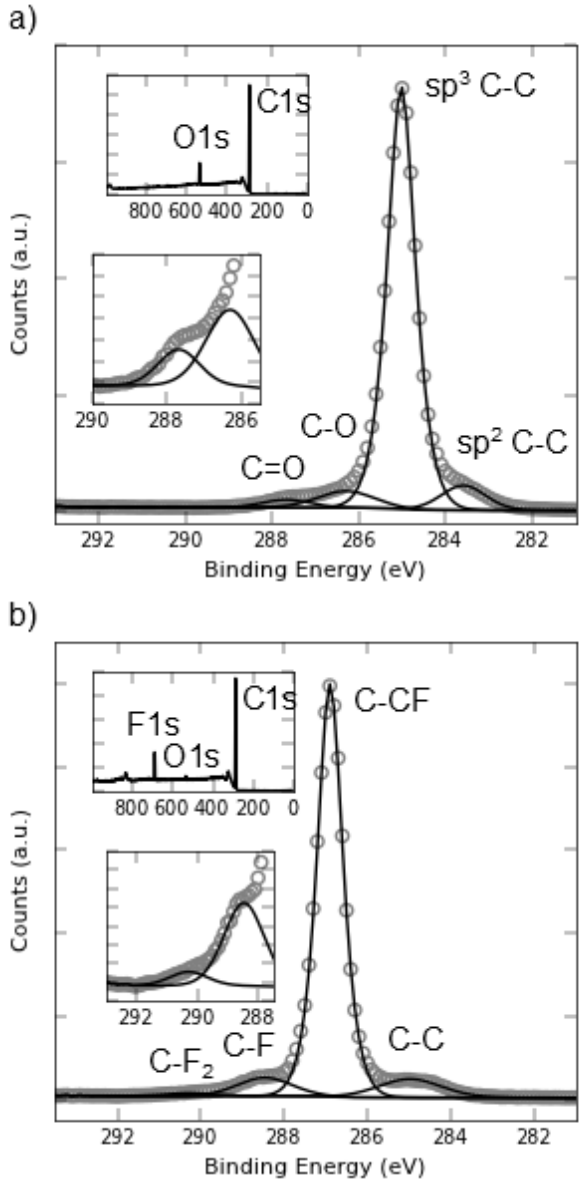

Figure 2. C 1s Core-level photoelectron spectra obtained for (a) oxygen- and (b) fluorine-terminated side of the same sample. Circles correspond to the measured values and solid lines to the fitted peaks. Top inset shows elemental composition of the respective termination. Bottom inset zooms into the higher binding energies to show the peaks more clearly.

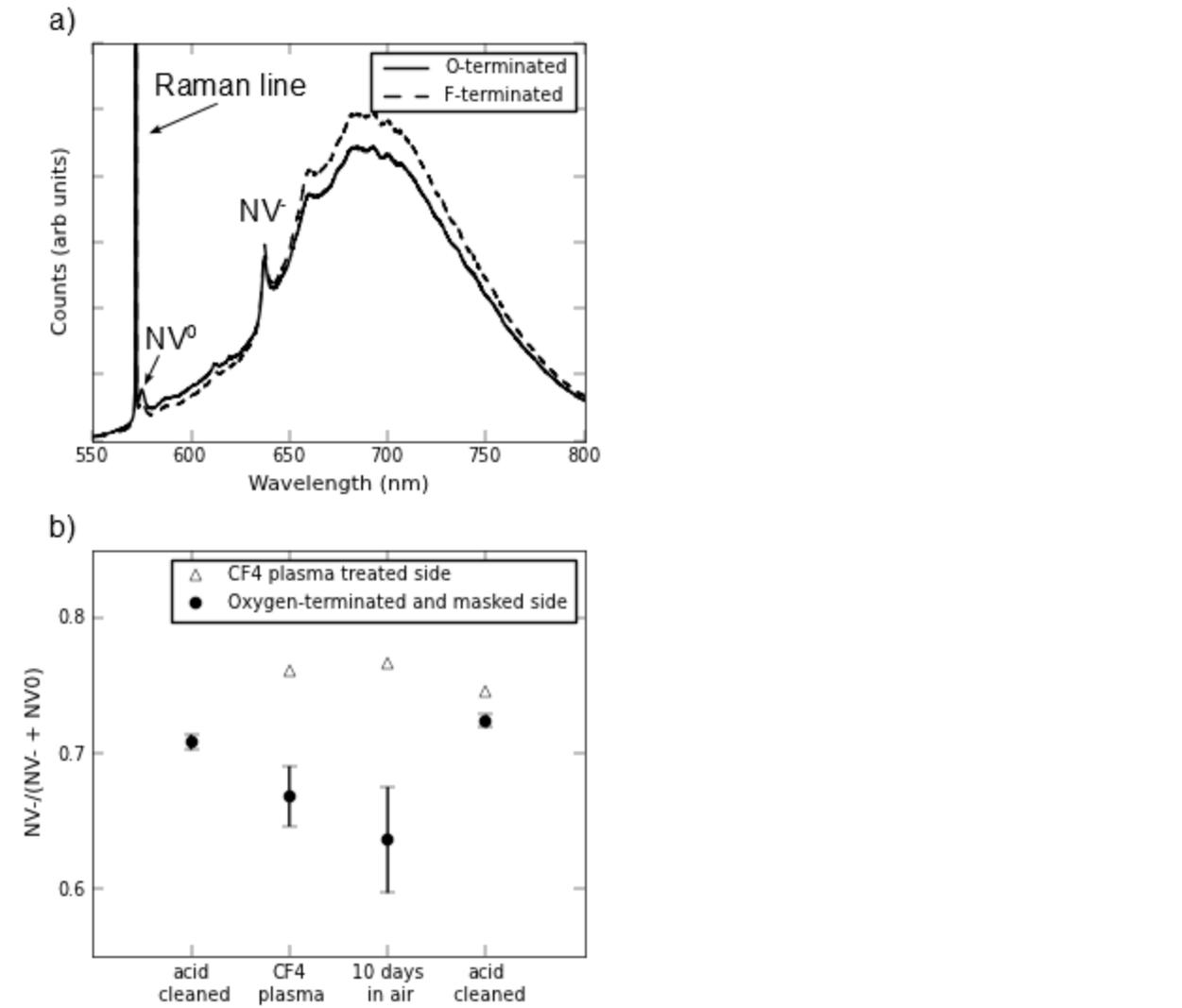

Figure 3. (a) Representative photoluminescence spectra of the oxygen- and fluorine-terminated side of the sample upon excitation at 532 nm. (b) The ratio of $NV^-$ ZPL area to the sum of $NV^-$ and $NV^0$ ZPL area of the same sample after 4 subsequent steps: 1. acid cleaned, 2. half protected from (circular markers) and exposed to (triangular markers) $CF_4$ plasma, 3. exposed to air for 10 days, and 4. re-oxygenated and cleaned with boiling triacid.

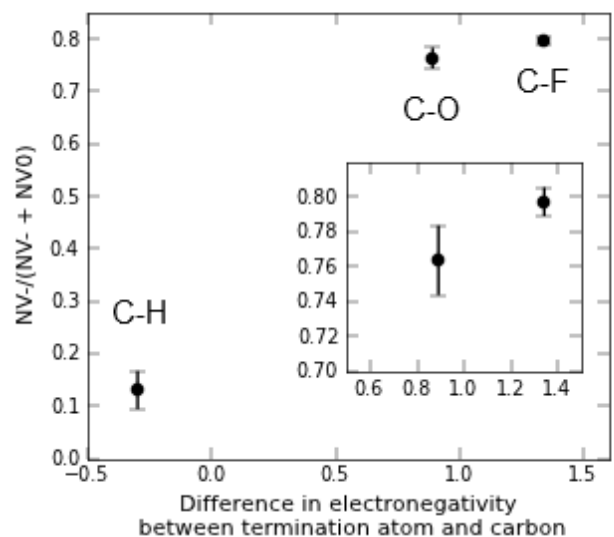

Figure 4. Ratio of NV$^-$ ZPL area to the sum of NV$^-$ and NV$^0$ ZPL area for hydrogen-, oxygen-, and fluorine-terminated surfaces. Inset: zoom in for O- and F-termination.